\documentclass[referee]{raa}           
\usepackage{graphicx,times}
\usepackage{natbib}
\usepackage{amssymb,amsmath}
\bibpunct{(}{)}{;}{a}{}{,}
\usepackage{pdflscape}
\usepackage{afterpage}
\usepackage{rotating}
\usepackage[a4paper=true,pagebackref=true]{hyperref}
\hypersetup{pdftitle = The title of my PDF, pdfauthor = My name, pdfsubject= The subject, pdfkeywords = keyword1 keyword2 keyword3} 
\hypersetup{colorlinks = true, linkcolor = green, anchorcolor = red, citecolor = blue, filecolor = red, pagecolor = red, urlcolor = red}

\begin{document}

   \title{Broad-band spectral study of LMXB black hole candidate 4U 1957+11 with \emph{NuSTAR}
}

 \volnopage{ {\bf 20XX} Vol.\ {\bf X} No. {\bf XX}, 000--000}
   \setcounter{page}{1}

   \author{Prince Sharma\inst{1*}, Rahul Sharma\inst{1,4}, Chetana Jain\inst{2*}, Gulab C. Dewangan\inst{3}, and Anjan Dutta\inst{1}}

   \institute{Department of Physics and Astrophysics, University of Delhi, Delhi 110007, India; {\it princerajsharma31@gmail.com}\\
\and
Hansraj College, University of Delhi, Delhi 110007, India; {\it chetanajain11@gmail.com}\\
\and
Inter-University Centre for Astronomy and Astrophysics (IUCAA), Pune, 411007, India\\
\and
Indian Institute of Science Education and Research (IISER) Mohali, Punjab, 140306, India\\
\vs \no
{\small Received 20XX Month Day; accepted 20XX Month Day}
}

\abstract{We present here the results of broadband spectral analysis of low-mass X-ray binary and a black hole candidate 4U 1957+11. The source was observed nine times with the Nuclear Spectroscopic Telescope Array (\emph{NuSTAR}) between 2018 September and 2019 November. During these observations, the spectral state of 4U 1957+11 evolved marginally. The disc dominant spectra are well described with a hot, multicolour disc blackbody with disc temperature varying in the range $kT_{\rm in} \sim$ 1.35--1.86 keV and a non-thermal component having a steep slope ($\Gamma =$ 2--3). A broad Fe emission line feature (5--8 keV) was observed in the spectra of all the observations. The relativistic disc model was used to study the effect of distance, inclination, and the black hole mass on its spin. Simulations indicate a higher spin for smaller distances and lower black hole mass. At smaller distances and higher mass, spin is maximum and almost independent of the distance. An inverse correlation exists between the spin and the spectral hardening factor for all the cases. The system prefers a moderate spin of about 0.85 for black hole masses between 4--6 M$_{\sun}$ for a 7 kpc distance.
}

   \authorrunning{P. Sharma et al. }            
   \titlerunning{Broad-band spectral study of 4U 1957+11}  
   \maketitle

%

\section{Introduction}           
\label{sect:intro}

4U 1957+11 is one of the few known persistent X-ray binary sources with a soft emission of $\sim$ 20--70 mCrab (2--12 keV). It was discovered with \emph{Uhuru} in 1973, during the scans of Aquila region \citep{Giacconi1974}. It is known to exist in the soft, disc-dominated spectral state for the past 45 years. \citet{Wijnands2002} reported a very weak X-ray variability (1--2 per cent root-mean square amplitude) and variation in the luminosity by a factor of about 4 on a time scale of months to years. Its optical counterpart, V1408 Aql was identified by \citet{Margon1978} with an orbital period of $9.329 \pm 0.011$ hr \citep{Thorstensen1987,Bayless2011}.

Independent studies of the source by \citet{White1984} and \citet{Schulz1989} have put the source close to several Black Hole Candidates (BHCs) in their high spectral states in the X-ray colour-colour diagram. The X-ray continuum of 4U 1957+11 has been described with various model components \citep{Nowak1999}. The interpretations from the different models, however, have led to different opinions on the nature of the compact object. Based on the Comptonization model, \citet{Singh1994} inferred the presence of a black hole (BH). However, the authors did not rule out the possibility of it being a weakly magnetised accreting neutron star (NS) and estimated a surface magnetic field strength of $\sim 10^{8}$ G. For the same data, \citet{Ricci1995} suggested the possibility of a BH as the primary. 

The \emph{Ginga} observed 4U 1957+11 in an unusually soft state and a two-component model comprising a disc blackbody component and a non-thermal power law component described the spectrum very well \citep{Yaqoob1993}. The authors used the derived values of the disc temperature ($kT_{\rm in}$) $\sim 1.5$ keV and inner disc radius, $r_{\rm in} \sqrt{\rm cos\theta} \sim 2$ km ( for d = 7 kpc ) to establish its similarity with neutron star low-mass X-ray binaries (NS-LMXBs). These results, however, depended strictly on the assumed distance and disc inclination. A detailed study of 4U 1957+11 with the \emph{ASCA}, \emph{RXTE}, and \emph{ROSAT} has given results biased towards a BH primary \citep{Nowak1999}. 4U 1957+11 is also known to show the presence of a hard-tail and mild X-ray variability during the high flux states with the increasing intensity \citep{Yaqoob1993,Wijnands2002,Bayless2011}. 

Observations of 4U 1957+11 with \emph{RXTE}/ASM has indicated long term periodicities of 117, 235, and 352 days. These have been attributed to be due to the precessing warped accretion disc \citep{Nowak1999}. The radiation-driven warping of disc shows that the system does not possess the critical binary radius to sustain a precessing disc \citep{Ogilvie2001}. However, it is proposed that the long-term X-ray variability can arise from the changes in the mass accretion rate through the disc with viscosity driven instability, as X-ray radiation from the inner region irradiates the disc \citep[][ and references therein]{Ogilvie2001,Wijnands2002}. 

The spectrum of the source has often been described with a multi-temperature blackbody ($kT_{\rm in}$= 1.2--1.8 keV) along with a significant contribution from a power law component and, rarely with, an emission feature around 6.5 keV \citep{Yaqoob1993,Ricci1995,Wijnands2002,Nowak2008}. The high resolution studies of 4U 1957+11 with \emph{Chandra}, \emph{RXTE} and \emph{XMM-Newton} are consistent with a low column density $N_{\rm H} \approx$ (1--2) $\times 10^{21}$ cm$^{-2}$ and a clean spectrum, free from any features. The Fe emission/absorption lines or/and Fe edge do not appear strongly or are absent in the disk spectra, making it an ideal system for probing the disc atmosphere and geometry of the system \citep{Nowak2008,Nowak2012}.

Since its discovery, 4U 1957+11 has never gone into quiescence, therefore an estimate of the mass function of the system is uncertain till date \citep{Casares2007}. The lack of eclipses and any modulations in the X-ray light curve rules out the possibility of inclination greater than $65^\circ$ \citep{Gomez2015}. \citet{Maitra2014} reported a high-inclination angle of $\sim 78^\circ$ and a distance of 5--10 kpc for a BH primary with mass $<$ 10 M$_{\sun}$ from their \emph{Swift}/XRT analysis. \citet{Gomez2015} have proposed a 3 M$_{\sun}$ BH ($< 6.2$ M$_{\sun}$) with a 1 M$_{\sun}$ companion and an inclination, $i = 12.75^\circ$ using a model that assumes the orbital modulation is caused by the irradiated face of the secondary star with its evolving orbital phase \citep{Thorstensen1987,Bayless2011}.
The joint \emph{Chandra}, \emph{RXTE} and \emph{XMM-Newton} spectral fits using the relativistic accretion disc model, suggests that 4U 1957+11 harbours either a 3 M$_{\sun}$ primary at a distance of 10 kpc with a moderate spin,  $a \approx$ 0.83 or a rapidly spinning ( $a \approx$ 1), 16 M$_{\sun}$ BH at 22 kpc. However, there exists degeneracies for the distance, mass and accretion rate \citep{Nowak2008}. From the \emph{Suzaku} observations, \citet{Nowak2012} estimated the spin parameter of  $a \gtrsim$ 0.9, for a 3 M$_{\sun}$ BH at a distance of 10 kpc ($i$ fixed to $75^\circ$) with the Comptonized disc as well as a relativistic disc model.

Generally, NSs exhibiting a soft spectra and a low root-mean square amplitude of variability show relatively high rates of Type I bursting activity. Therefore, absence of Type I X-ray bursts in 4U 1957+11 is a suggestive evidence for presence of a BH primary in this system \citep{Wijnands2002,Maccarone2020}. Although there have been several detailed studies exploring the nature of the compact object in 4U 1957+11, no study has been compelling and conclusive. The uncertain physical parameters and poor description of the system attributes make it difficult to classify the compact object in the system. We have performed a conservative analysis for the case of a BH primary and have discussed the broad-band spectral characteristics for the system.

\section{Observations and data analysis}
\label{sect:Obs}

\begin{table}
	\centering
	\caption{Log of the \emph{NuSTAR} observations used in this work.}
	\label{tab:obslog}
	\begin{tabular*}{\textwidth}{@{\extracolsep{\fill}} cccccccc}
	\hline
		Observation & Observation ID & Observation Date & MJD & Exposure$^{a}$ & Energy & Count Rate$^{b}$\\
		Epoch & & dd-mm-yyyy hh:mm:ss & (d) & (ks) & (keV) & (c/s)\\ 
		\hline
		1 & 30402011002 & 16-09-2018 07:16:09 & 58377.31  & 74.39 & 3.0--20.0 & $7.56 \pm 0.01$\\[0.5ex]
		2 & 30402011004 & 13-03-2019 10:01:09 & 58555.41  & 67.01 & 3.0--45.0 & $17.24 \pm 0.02$\\[0.5ex]
		3 & 30502007002 & 29-04-2019 20:41:09 & 58602.86  & 41.43 & 3.0--40.0 & $11.41 \pm 0.02$\\[0.5ex]
		4 & 30402011006 & 15-05-2019 12:01:09 & 58618.50  & 74.04 & 3.0--50.0 & $23.00 \pm 0.02$\\[0.5ex]
		5 & 30502007004 & 04-06-2019 19:56:09 & 58638.83  & 40.11 & 3.0--50.0 & $22.23 \pm 0.02$ \\[0.5ex]
		6 & 30502007006 & 19-07-2019 06:11:09 & 58683.26 & 36.76 & 3.0--50.0 & $27.55 \pm 0.03$ \\[0.5ex]
		7 & 30502007008 & 10-09-2019 01:36:09 & 58736.07 & 20.25 & 3.0--50.0 & $31.37 \pm 0.04 $ \\[0.5ex]
		8 & 30502007010 & 20-10-2019 12:51:09 & 58776.54 & 39.23 & 3.0--25.0  & $19.43 \pm 0.02 $\\[0.5ex]
		9 & 30502007012 & 30-11-2019 20:56:09 & 58817.87 & 40.96 & 3.0--20.0  & $11.20 \pm 0.02 $\\[0.7ex]
	\hline
	\multicolumn{7}{l}{\textit{Notes.} $^{a}$Exposure for the combined and clean \emph{NuSTAR} FPMA and FPMB spectra after good time filtering.}\\
    \multicolumn{7}{l}{$^{b}$Average count rate for the corresponding energy range.}\\
	\end{tabular*}
\end{table}

\begin{figure}
\centering
	\includegraphics[width=0.8\columnwidth]{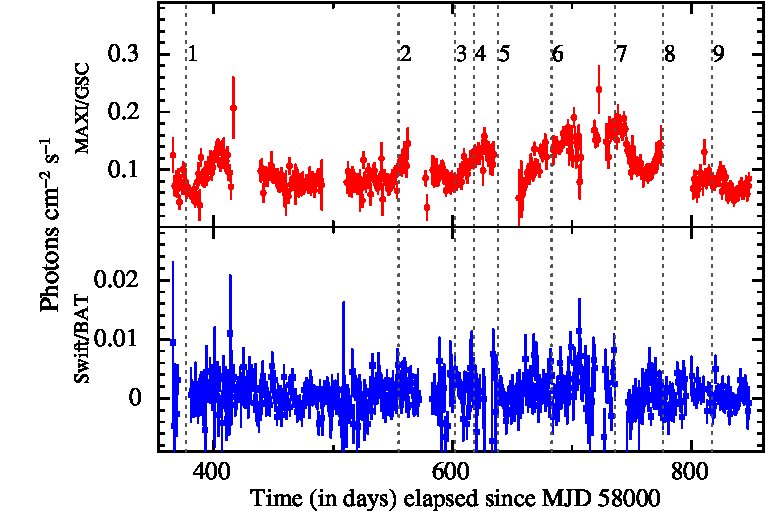}
    \caption{The long term \emph{MAXI}/GSC (2--4 keV, top panel) and \emph{Swift}/BAT (15--50 keV, bottom panel) daily-averaged light curves of 4U 1957+11. The vertical lines mark the epochs of nine \emph{NuSTAR} observations that have been used in this work.} 
    \label{fig:longlc}
\end{figure}

Nuclear Spectroscopic Telescope Array (\emph{NuSTAR}) is the NASA's first high-energy space mission consisting of advanced X-ray focusing telescopes that provide good sensitivity above 10 keV for the imaging purposes \citep{Harrison2013}. Its payload comprises two identical detectors, known as focal plane module A and B (FPMA and FPMB) fixed at the focus of each of the two co-aligned, grazing incidence telescopes with a field of view of 10 arcmin$^2$ and an energy resolution (full-width half-maximum) of 400 eV at 10 keV. It observed 4U 1957+11 nine times between 2018 September and 2019 November, for a total exposure of $\sim 217$ ks per module. We have analysed all the \emph{NuSTAR} observations for our broad-band spectral study of the source that comprises \emph{NuSTAR} Guest Observer (GO) cycle 1 and joint \emph{NuSTAR/NICER} GO cycle 1. The details of the observations are listed in Table~\ref{tab:obslog}. Figure~\ref{fig:longlc} shows the daily-averaged, \emph{Swift}/BAT (15--50 keV)\footnote{https://swift.gsfc.nasa.gov/results/transients/weak/4U1957p115/} \citep{Krimm2013} and \emph{MAXI}/GSC (2--4 keV)\footnote{http://maxi.riken.jp/pubdata/v7l/J1959+117/index.html} \citep{Matsuoka2009} long-term light curves starting from the 2018 September. The dotted vertical lines mark the epochs of the nine observations used in this work. The X-ray emission varied during the observations. The X-ray flux during the epochs 6 and 7 was relatively higher as compared to the flux during the other epochs.

\subsection{NuSTAR Spectra}
\label{sec:nu} 
We have reduced the \emph{NuSTAR} data with the \textsc{nustardas}\_01Apr20\_v1.9.2 and \textsc{caldb} v20200912. We used the task \textsc{nupipeline} that applies the standard screening and calibration to the raw event files and produces level 2 files. The spectra, light curves and response files were generated by processing the level 2 files with \textsc{nuproducts}. The source spectra and light curves were extracted from a circular region of radius 100 arcsec centred on the source for both FPMA and FPMB. We used a circular region of the same size, away from the source, to extract the background spectra and light curves. We co-added the two source spectra, background spectra and response files from FPMA and FPMB for each observation by using \textsc{addspec} tool, to improve the efficiency and statistics. Table~\ref{tab:obslog} reports the final exposure of spectra. Finally, all the nine spectra were re-binned to contain a minimum of 20 counts in each energy bin.  

The emission spectrum of the source evolved through the nine observation epochs and source was detected above the background differently for different epochs. The upper energy bounds were decided based on the detection of source emission above background, while the lower bound of 3 keV ascribed to the calibration uncertainties in the responses near the lower instrumental band-pass. All the \emph{NuSTAR} spectra for 4U 1957+11 were modelled in the energy bands as mentioned in the Table~\ref{tab:obslog}.  

\section{Light curves and variability}

\begin{figure}
\centering
	\includegraphics[width=0.8\columnwidth]{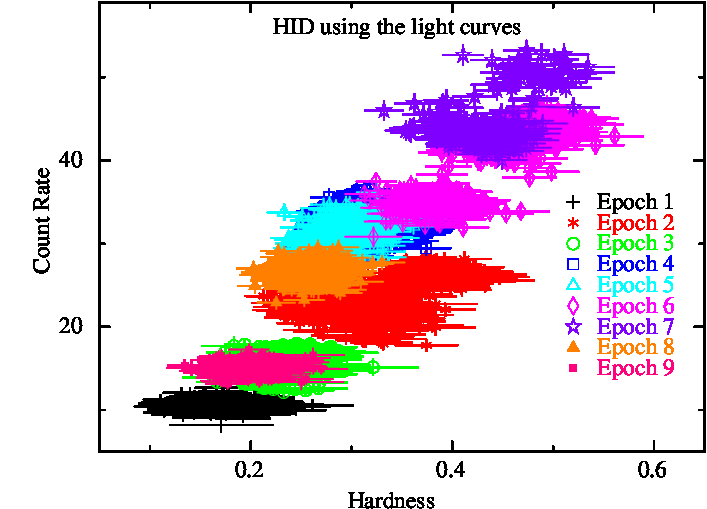}
	 \caption{The hardness-intensity plot for the nine \emph{NuSTAR} observations of 4U 1957+11. The soft X-ray band is taken as 3--7 keV and hard X-ray band is in the range 7--20 keV.}
    \label{fig:hid1}
    \end{figure} 

\begin{figure}
	\includegraphics[width=0.5\columnwidth]{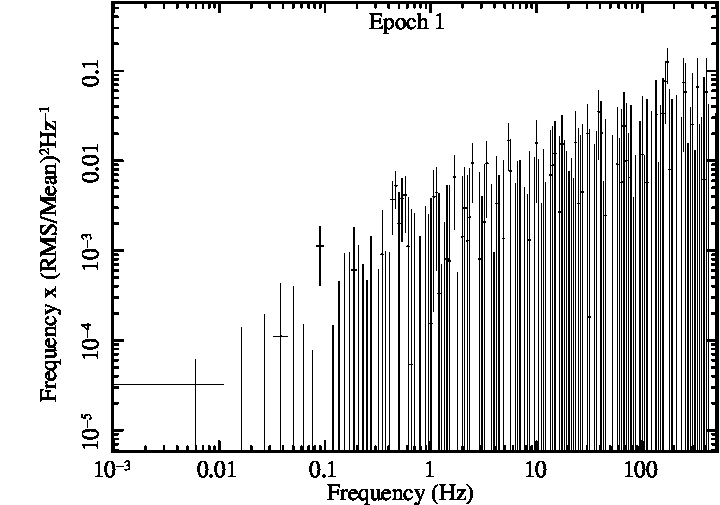}
	\includegraphics[width=0.5\columnwidth]{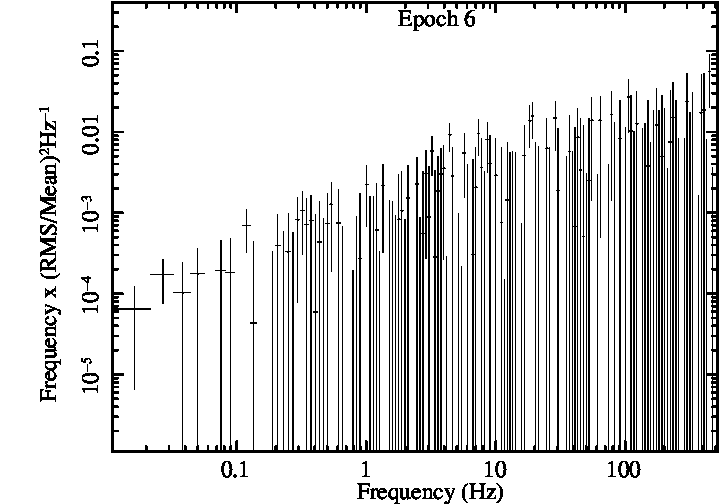}
    \caption {The 3--20 keV power density spectra of 4U 1957+11 for the first (\textit{left}) and sixth (\textit{right}) observation. The PDS are plotted in the $\nu$P($\nu$) representation.}
    \label{fig:pds}
    \end{figure}

We generated the background subtracted light curves of 4U 1957+11 from all the nine observations. We did not find any Type I X-ray burst in the entire data. This provides a support against the case of a NS primary \citep{Maccarone2020}. As seen in Figure~\ref{fig:longlc}, X-ray flux changed during the observing span. In the soft X-ray band ($< 7$ keV), the average count rate varied by a factor of $\sim$ 3.5. It varied by a factor of $\sim$ 11 in the hard X-ray band ($> 7$ keV). 
 
Figure~\ref{fig:hid1} shows the hardness intensity plot generated from the background subtracted light curves. The hardness varied from $\sim$ 0.17 during the first epoch to about $\sim$ 0.43 during the sixth and seventh epochs and back to $\sim$ 0.20 during the last observation. Combined with Figure~\ref{fig:longlc}, it appears that amongst all the observations, epochs six and seven correspond to a marginally hard spectral state along with a comparatively higher count rate. 
 
We also generated power density spectra (PDS) by using the \textsc{hendrics} 5.0 package \citep{Bachetti2015}. We used the barycenter corrected light curves in the energy range of 3--20 keV to generate the PDS. We used the root-mean-square normalization and rebinned the PDS geometrically by a factor of 0.01. The PDS from all the observations showed insignificant variability for frequencies up to 512 Hz. Figure~\ref{fig:pds} shows the PDS for the first and sixth epochs.

\begin{figure}
	\centering 
	\includegraphics[width=0.8\columnwidth]{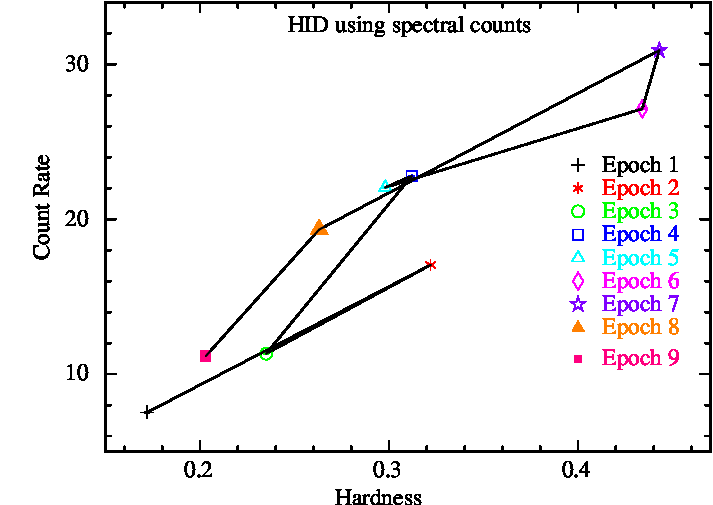}
	 \caption{The hardness intensity diagram (HID) for 4U 1957+11 generated from the \emph{NuSTAR} observations by using the average spectral counts. The soft X-ray band is taken as 3--7 keV and hard X-ray band is in the range 7--20 keV.}
    \label{fig:hid2}
    \end{figure}

As a cross-check for changes in the spectral states (if any), we created the hardness intensity diagram (HID) by using the average spectral counts (Figure~\ref{fig:hid2}). We took 3--7 keV as the soft energy range and 7--20 keV as the hard energy range. It is clear that during the \emph{NuSTAR} observations, 4U 1957+11 started from a low hardness during the first observation. It became significantly harder and brighter and occupied the top-right corner of the HID during the sixth and seventh observations. Its brightness decreased thereafter.
 
The evolution of the source emission is not very significant, however, for our analysis, we have categorized the observations in groups based on the spectral components required to model the spectra.

\section{Specral analysis and results}

\begin{figure}
\centering
	\includegraphics[width=0.8\columnwidth]{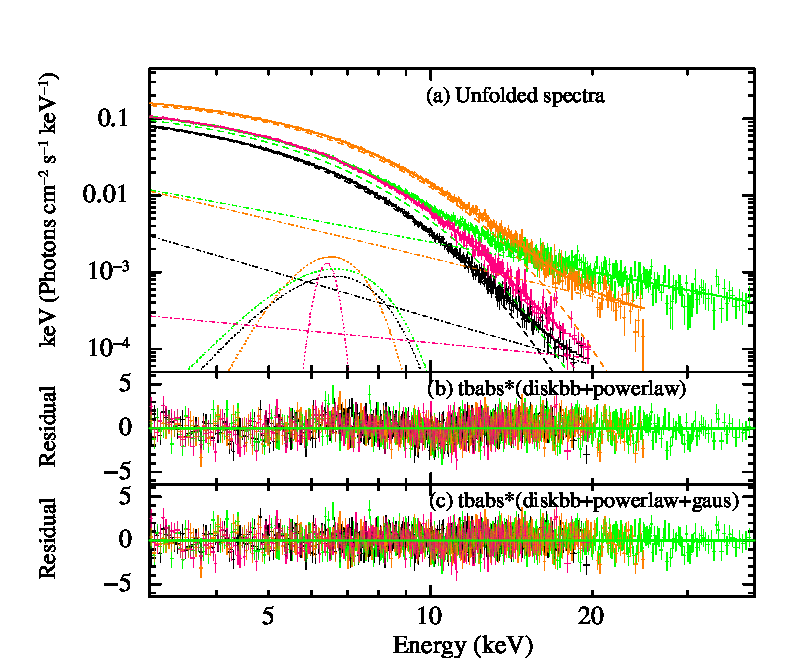}
	\caption{The unfolded \emph{NuSTAR} spectra and model components for the epochs 1, 3, 8, and 9. (a) Best fit spectra with the model comprising \texttt{diskbb} and \texttt{powerlaw}. (b) Residuals for the model without a Gaussian component. (c) After adding the Gaussian component. Colour scheme is same as Figure~\ref{fig:hid1}. The dash, dash-dot, and dot-dot lines represent \texttt{diskbb}, \texttt{powerlaw}, and \texttt{gaus} components, respectively. The spectra have been rebinned for clear representation.}
	\label{fig:dbpoga_1389}
	\end{figure}
	
\begin{figure}
\centering
	\includegraphics[width=0.8\columnwidth]{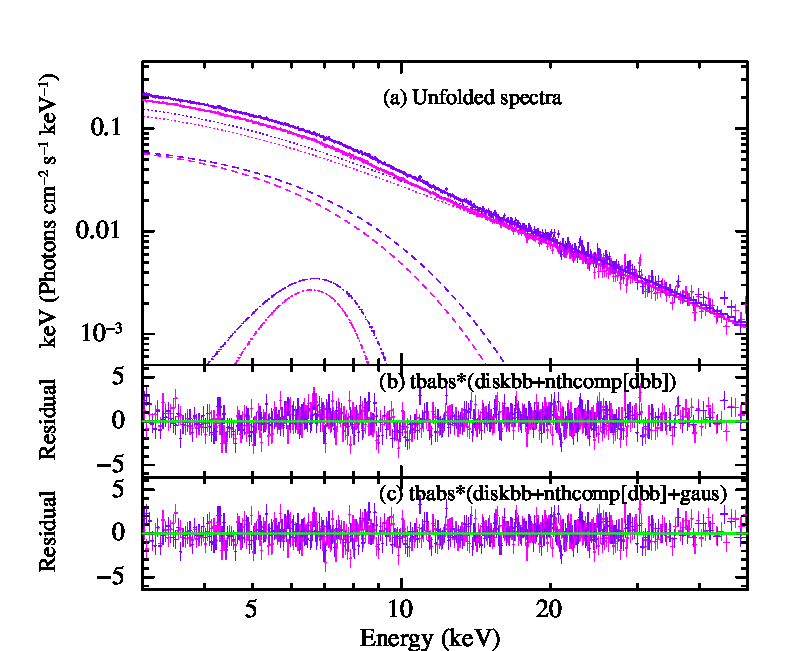}
	\caption{(a) The unfolded \emph{NuSTAR} spectra and model components for the epochs 6 and 7 modelled with \texttt{tbabs*(diskbb+nthcomp+gaus}). (b) Residuals for the model without a Gaussian component. (c) After adding the Gaussian component. Colour scheme is same as Figure~\ref{fig:hid1}. The dash, dash-dot, and dot-dot lines represent \texttt{diskbb}, \texttt{nthcomp}, and \texttt{gaus} components, respectively. The spectra have been rebinned for clear representation.}
	\label{fig:dbnth67}
	\end{figure}

\begin{table}
\centering
	\caption{Best fit spectral parameters for the \emph{NuSTAR} spectra from the nine epochs. All the parameter uncertainties are given at 90 per cent confidence level.}
	\label{tab:spectral}
	\scalebox{0.65}{     
	\begin{tabular}{ccccccccccc}
	\hline
    Component & Parameters & Epoch 1 & Epoch 2 & Epoch 3 & Epoch 4 & Epoch 5 & Epoch 6 & Epoch 7 & Epoch 8 & Epoch 9 \\
    \hline

    DISKBB & $kT_{\rm in}$ (keV) & $1.36 \pm 0.01$ & $1.56_{-0.02}^{+0.01} $ & $1.42 \pm 0.01$ & $1.65 \pm 0.01$ & $1.65 \pm 0.01$ & $1.63 \pm 0.07$ & $1.79^{+0.07}_{-0.12}$ & $1.63 \pm 0.01$ & $1.47 \pm 0.01$ \\[0.5ex]
     & $N_{\rm disc}$ & $10.35_{-0.28}^{+0.26}$ & $7.52_{-0.56}^{+0.38}$ & $10.46_{-0.26}^{+0.30}$ & $8.93_{-0.20}^{+0.22}$ & $9.72_{-0.19}^{+0.29}$ & $3.85^{+0.85}_{-0.60}$ & $2.85^{+0.83}_{-0.49}$ & $9.95_{-0.21}^{+0.19}$ & $10.30_{-0.17}^{+0.14}$\\[0.5ex]
     & $f_{\rm dbb} $  & $7.50 \pm 0.03$ & $ 9.57 \pm 0.03 $ & $ 9.23 \pm 0.04 $ & $ 14.16 \pm 0.03 $ & $ 15.62 \pm 0.04 $ & $5.80 \pm 0.05$ & $6.24 \pm 0.06$ & $15.10 \pm 0.04$ & $10.34 \pm 0.04$\\[2ex]

    POWERLAW  & $\Gamma$ & $3.04_{-0.44}^{+0.28}$ & - & $2.31 \pm 0.08$ & - & - & - & - & $2.67_{-0.17}^{+0.13}$ & $1.68_{-1.86}^{+1.52}$ \\[0.5ex]
    & Norm ($10^{-2}$) & $2.82_{-2.02}^{+3.20}$ & - & $5.04_{-1.02}^{+1.31}$ & - & - & - & - & $7.23_{-2.81}^{+3.47}$ & $0.06_{-0.05}^{+1.24}$ \\[0.5ex]
    & $f_{\rm pl}$ & $4.80_{-2.60}^{+3.76}$ & - & $4.71_{-0.78}^{+0.92}$ & - & - & - & - & $8.04_{-1.86}^{+2.13}$ & $0.11_{-0.02}^{+0.19}$ \\[2ex]
    
    NTHCOMP  & $\Gamma$ & - & $2.41_{-0.10}^{+0.21}$ & - & $2.60_{-0.09}^{+0.05}$ & $2.43_{-0.21}^{+0.09}$ & $2.80_{-0.10}^{+0.02}$ & $2.81_{-0.22}^{+0.07}$ & - & - \\[0.5ex]
    & $kT_{\rm e}$ (keV) & - & $> 6.49$ & - & $> 1.01 $ & $> 1.01$ & $> 1.01$ & $> 1.01 $ & - & - \\[0.5ex]
    & $kT_{\rm seed}$ (keV) & - & $< 0.55$ & - & $0.63 \pm 0.04$ & $< 0.63$ & $1.15_{-0.07}^{+0.09}$ & $1.12_{-0.12}^{+0.09}$ & - & - \\[0.5ex]
    & Norm ($10^{-2}$) & - & $6.23_{-1.62}^{+10.37}$ & - & $3.00_{-0.38}^{+0.45}$ & $2.48_{-0.35}^{+12.28}$ & $17.13_{-1.82}^{+1.03}$ & $20.48_{-2.08}^{+1.62}$ & - & - \\[0.5ex]
    & $f_{\rm nth}$ & - & $5.98 \pm 0.13$ & - & $5.11 \pm 0.11$ & $3.48 \pm 0.14$ & $ 18.65\pm0.29 $ & $21.56 \pm 0.34 $ & - & -\\[2ex]
    
    GAUS & $E_{\rm line}$ (keV) & $ < 6.64$ & $ < 6.49$ & $< 6.56$ & $< 6.52$ & $< 6.49$ & $<6.56$ & $< 6.60$ & $< 6.56$ & $< 6.52$ \\[0.5ex]
    & $\sigma$ (keV) & $1.26_{-0.21}^{+0.16}$ & $1.02 \pm 0.22$ & $1.30_{-0.28}^{+0.22}$ & $1.18_{-0.21}^{+0.20}$ & $0.46_{-0.22}^{+0.68}$ & $1.10_{-0.31}^{+0.25}$ & $1.38_{-0.27}^{+0.25}$ & $0.90_{-0.28}^{+0.26}$ & $0.25_{-0.12}^{+0.18}$ \\[0.5ex]
    & Norm ($10^{-4}$) & $4.29_{-1.33}^{+1.14}$ & $5.43_{-2.18}^{+3.00}$ & $5.52_{-2.06}^{+2.30}$ & $9.08_{-2.53}^{+2.85}$ & $2.69_{-1.17}^{+3.70}$ & $11.45_{-3.86}^{+4.51}$ & $18.32_{-6.80}^{+7.49}$ & $5.50_{-2.06}^{+2.29}$ & $1.26_{-0.46}^{+0.57}$ \\[0.5ex]
    & EW (eV) & $134_{-80}^{+101}$ & $72_{-53}^{+61}$ & $125_{-113}^{+103}$ & $64_{-28}^{+91}$ & $27_{-15}^{+16}$ & $93_{-49}^{+150} $ & $< 436$ & $61_{-32}^{+35}$ & $36_{-14}^{+2}$ \\[0.5ex]
    & $f_{\rm gau}$ & $0.44 \pm 0.06$ & $ 0.56 \pm 0.09 $ & $0.57 \pm 0.11$ & $0.93 \pm 0.12 $ & $0.28 \pm 0.08$ & $1.17 \pm 0.17$ & $1.88 \pm 0.31$ & $0.56 \pm 0.12$ & $0.13 \pm 0.04$ \\[2ex]
    
    &$f_{\rm Tot}$ & $12.35 \pm 0.03$ & $15.61 \pm 0.04 $ & $13.98 \pm 0.03$ & $19.36 \pm 0.04$ & $19.13 \pm 0.05$ & $24.57 \pm 0.04$ & $27.98 \pm 0.06$ & $23.20 \pm 0.04$ & $10.47 \pm 0.02$\\[1ex]
    \hline
    & $\chi^2$/d.o.f & $294.4/306$ & $774/692$ & $528.2/486$ & $852.4/717$ & $569.3/567$ & $734.1/740$ & $650.1/658$ & $408.2/400$ & $315/308$ \\
    \hline
	\multicolumn{11}{l}{\textit{Note.} $f_{\rm dbb} $, $f_{\rm pl}$, and $f_{\rm Tot}$ give the unabsorbed flux in the energy band 0.1--100 keV in units of $10^{-10}$ erg cm$^{-2}$ s$^{-1}$  .}\\
	\multicolumn{11}{l}{$f_{\rm gau}$ is the unabsorbed flux in the energy band 0.1--100.0 keV in units of  $10^{-11}$ erg cm$^{-2}$ s$^{-1}$.}\\
	\multicolumn{11}{l}{Normalization for the \texttt{diskbb} component is defined as $(R_{\rm in}(\rm km))^2/(D/{\rm 10\ kpc})^{2} {\rm {cos}\theta}$.}\\
	\end{tabular}}
\end{table}

We have used the spectral analysis package \textsc{xspec} v12.11.0k \citep{Arnaud1996}, distributed with the \textsc{heasoft} v6.27, to model the \emph{NuSTAR} spectra of 4U 1957+11 and have used the $\chi^2$ statistics as the test statistics. We have adopted the photoelectric cross-sections from \citet{Verner1996} and solar abundances from \citet{Wilms2000}. All the uncertainties, lower limits and upper limits are reported at 90 per cent confidence level. 

We started with the two-component model: a multi-temperature disc blackbody, \texttt{diskbb} \citep{Mitsuda1984}, to account for the emission from the accretion disc, and the thermal Comptonization component \texttt{nthcomp}, \citep{Zdziarski1996,Zycki1999} with low-energy and high-energy rollovers to model the hard emission. We set the seed photon source to the accretion disc. We also included the \texttt{tbabs} to account for the absorption due to the interstellar medium and tied it across all the spectra. The value of the column density was not constrained, when left free, due to the low column density in the direction of 4U 1957+11. Therefore, we fixed the $N_{\rm H}$ to 1.22$\times 10^{21}$ cm$^{-2}$ \citep{Maitra2014}. This is consistent with a value of 1.17$\times 10^{21}$ cm$^{-2}$ calculated from the HI 4 Pi survey \citep{HIPI}\footnote{https://heasarc.gsfc.nasa.gov/cgi-bin/Tools/w3nh/w3nh.pl}.

The model, \texttt{tbabs*(diskbb+nthcomp)} (M1, hereafter) provided a statistically poor fit with a $\chi^{2}$ of 5475.2 for 4895 degrees of freedom (d.o.f). Due to the weak emission during the first, eighth, and ninth epochs, the \texttt{nthcomp} parameters could not be constrained for these epochs. We used the simple \texttt{powerlaw} component along with the \texttt{diskbb} to model the spectra from these three epochs. Since the third epoch shows a similar spectral behaviour (Figure~\ref{fig:hid2}), therefore, its spectrum was also fit with the spectra from these three epochs. Initially, the model yielded a poor fit with a $\chi^2$ of 1746.4 for 1512 d.o.f and a broad feature between 5 and 8 keV, possibly due to the Fe K-line emission. We used the Gaussian component to model the excess around 6.5 keV in the four spectra and constrained its energy between 6.4 and 7 keV, leaving the other parameters free. The resulting model, \texttt{tbabs*(diskbb+powerlaw+gaus)} provided an acceptable fit with a $\chi^2$/d.o.f = 1595.8/1500. The best fit spectra and model components are shown in Figure~\ref{fig:dbpoga_1389}. The best fit spectral parameters for this model and the subsequent analysis are reported in the Table~\ref{tab:spectral}.

The model M1 provided a statistically acceptable fit with a $\chi^{2}$/d.o.f = 1464.4/1404 to the spectra from epochs 6 and 7. However, the broad emission feature was present in the residuals (Figure~\ref{fig:dbnth67}). The addition of Gaussian component to model the excess improved the fit by reducing $\chi^2$ by 80 for 6 additional parameters. The best fit returned the disc temperature of $1.63 \pm 0.07$ and $1.79^{+0.07}_{-0.12}$ keV along with very low normalization ($N_{\rm disc}$) $3.85^{+0.85}_{-0.60}$ and $2.85^{+0.83}_{-0.49}$ for epochs 6 and 7, respectively. However, the electron corona temperature could not be constrained and we obtained the lower limits of 1.01 keV.

\begin{figure}
\centering
	\includegraphics[width=0.8\columnwidth]{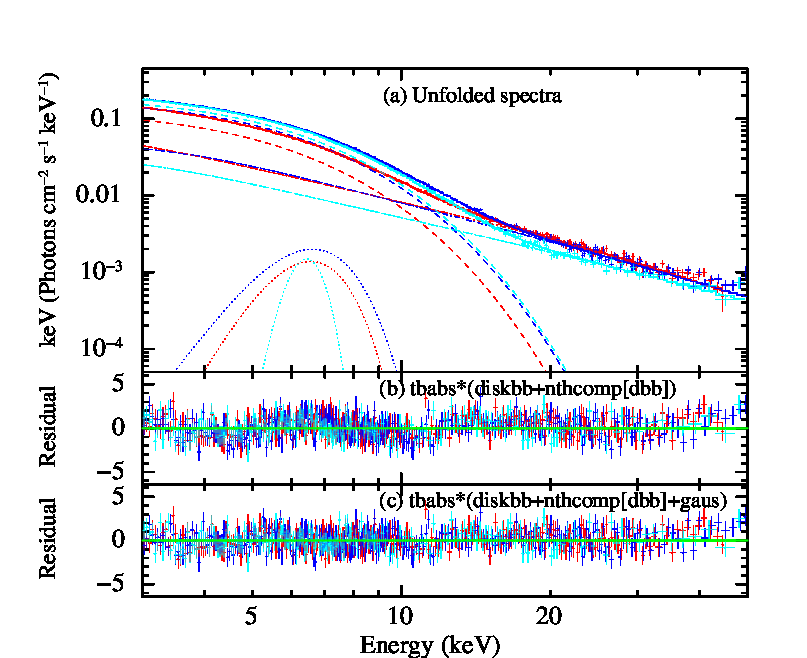}
	\caption{The unfolded \emph{NuSTAR} spectra and model components for the epochs 2, 4, and 5. (a) Best fit spectra with the model comprising \texttt{diskbb} and \texttt{nthcomp}. (b) Residuals for the model without a Gaussian component. (c) After adding the Gaussian component. Colour scheme is same as Figure~\ref{fig:hid1}. The dash, dash-dot, and dot-dot lines represent \texttt{diskbb}, \texttt{nthcomp}, and \texttt{gaus} components, respectively. The spectra have been rebinned for clear representation.}
	\label{fig:dbnthga245}
	\end{figure}

When used for the epochs 2, 4, and 5 the model M1 provided a statistically poor fit ($\chi^2$/d.o.f = 2328.5/1985) to the spectra due to the presence of a broad Fe emission feature between 5 and 8 keV. We added a Gaussian component to model this excess as before. The addition of Gaussian component improved the fit by reducing the $\chi^2$ by 132.8 for 9 additional parameters (Figure~\ref{fig:dbnthga245}). The final model, \texttt{tbabs*(diskbb+nthcomp+gaus)} resulted in high disc temperatures $kT_{\rm in}$ of $1.56^{+0.01}_{-0.02}$, $1.65 \pm 0.01$, and $1.65 \pm 0.01$ keV, with disc normalization of $7.53^{+0.38}_{-0.56}$, $8.93_{-0.20}^{+0.22}$, and $9.72_{-0.19}^{+0.29}$ for epochs 2, 4, and 5, respectively. We obtain the upper limits of 6.49, 6.52, and 6.49 keV for the emission line energy for epochs 2, 4, and 5, respectively.

We used the convolution model \texttt{cflux} and calculated unabsorbed flux in the 0.1--100keV band for the full model as well as the individual components.The net flux increased by a factor of 1.3 from epoch 1 to 2 before declining marginally during epoch 3. It then increased gradually and reached a maximum during the epoch 7 becoming 2.3 times the initial value before decreasing during the last two epoch. The fractional contribution of the thermal component to the net flux increased gradually from 0.61 to 0.82 between epochs 1 and 5. It then decreased to 0.23 during epochs 6 and 7 before increasing back to $\sim 1$ during the last epoch. The relatively high flux states exhibited a significant contribution from the hard emission component (\texttt{nthcomp}) accompanied by the reduced thermal contribution, compared to the low flux states.

\subsection{Relativistic disc model}
\begin{figure}
	\includegraphics[width=0.5\columnwidth]{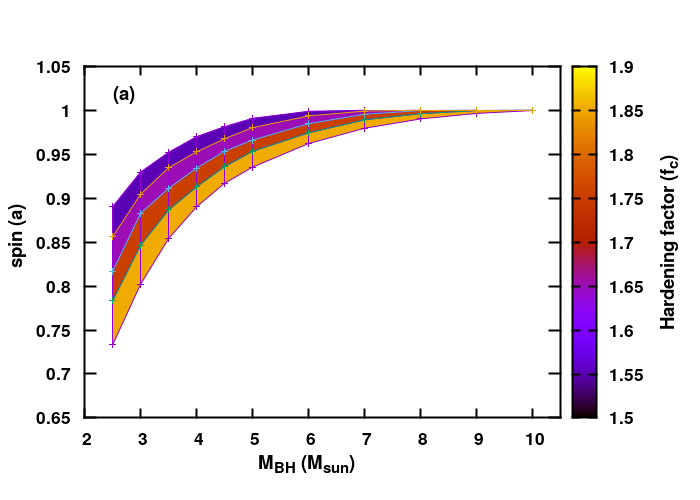}
	\includegraphics[width=0.5\columnwidth]{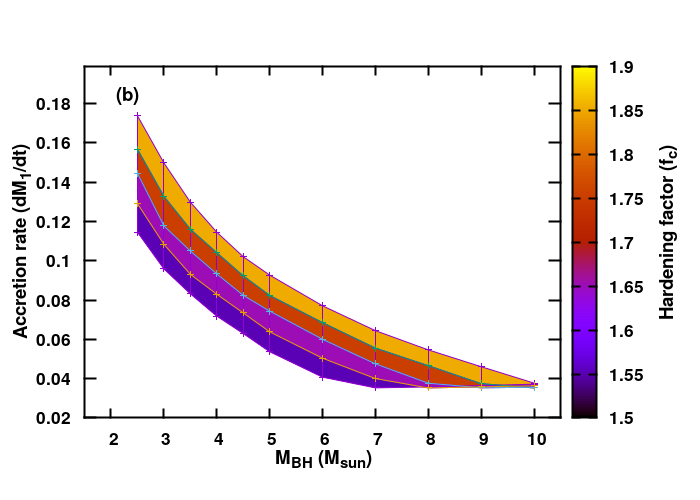}
        \includegraphics[width=0.5\columnwidth]{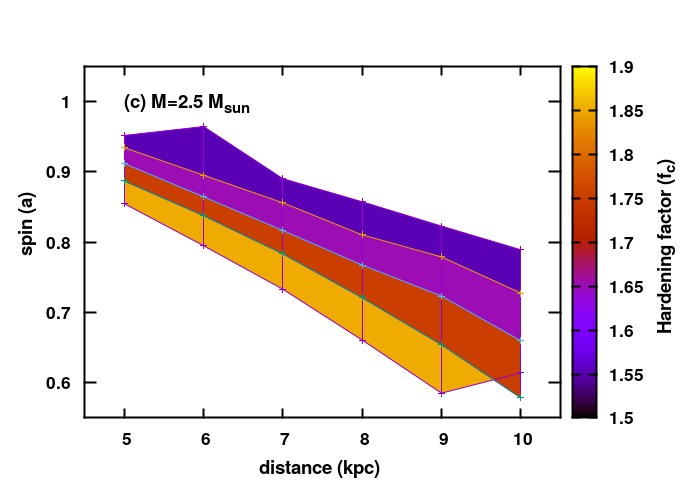}
	\includegraphics[width=0.5\columnwidth]{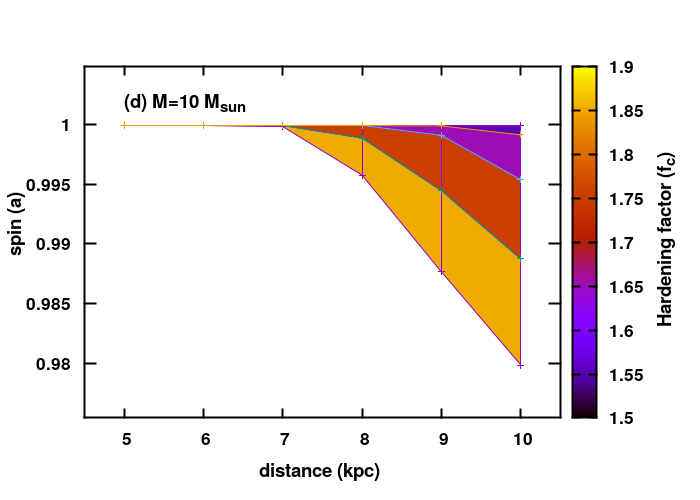}
        \includegraphics[width=0.5\columnwidth]{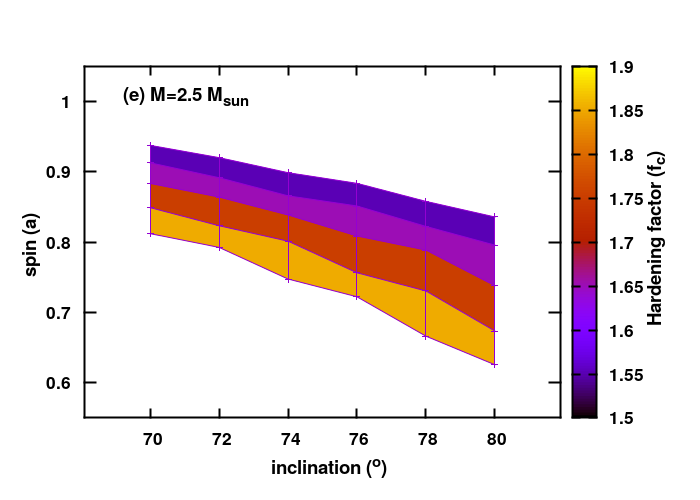}
	\includegraphics[width=0.5\columnwidth]{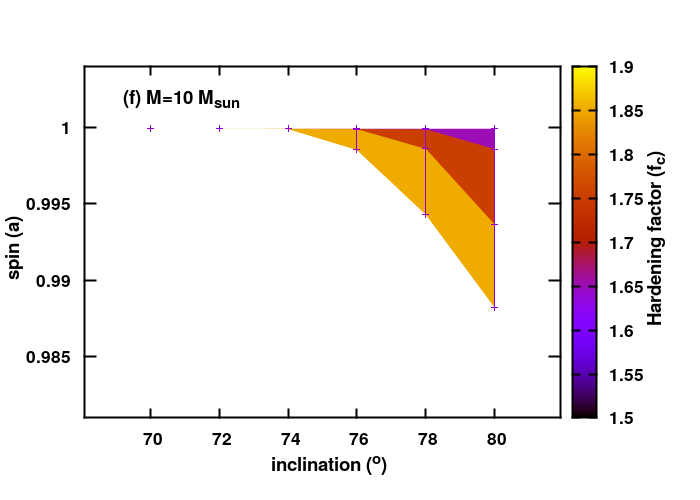}
	\caption{Heat maps from the best fit spectra from sixth and seventh epochs. A model \texttt{tbabs*ThComp*(kerrbb+gaus)} was used to generate these maps. The doublets of mass, distance, and inclination with spectral hardening factor are represented by coloured blocks. The `+' correspond to the best fit spin value for (a), (c), (d), (e), and (f). In (b), the `+' indicates the best fit value for the accretion rate.}
	\label{fig:mafc}
	\end{figure}

The previous studies of the source have used the relativistic, thin accretion disc model around a rotating BH, \texttt{kerrbb} \citep{Li2005} that incorporates the multi-temperature blackbody and considers the self-irradiation of the disc \citep{Nowak2008,Nowak2012}. It is a substitute for \texttt{diskbb} model with a total of eight parameters including mass of the BH ($M$), disc inclination ($i$), distance ($d$), ratio of the disc power produced by a torque at the inner edge to the power arising from accretion ($\eta$), spectral hardening factor ($f_{\rm c}$), and accretion rate ($\dot{M}$). Generally, the physical parameters such as $M$, $d$, and $i$ are known and are fixed during the spectral fitting. But in the case of 4U 1957+11, the constraint on mass and inclination is not well known. The earlier studies have provided the bounds on the distance of the binary system as the closest possible distance greater than 10 kpc \citep{Nowak2008,Maitra2014} and a maximum limit of 50 kpc, for the system to reside in the galaxy \citep{Russell2010,Gomez2015}. In a recent study, \citet{Maccarone2020} reported the most probable source distance of 7 kpc based on the posterior probability with 3 kpc and 15 kpc as the $5^{\rm th}$ and $95^{\rm th}$ percentile of distribution by using the \emph{Gaia} parallax measurement.

For completeness of this work, we have tried to fit the spectra to estimate the spin and accretion rate of the system. We modelled the spectra with the \texttt{kerrbb} model and the convolving component \texttt{ThComp}. The \texttt{ThComp} is described by the photon index $\Gamma$, electron temperature $kT_{\rm e}$, scattered fraction $f_{\rm sc}$ and assumes a spherical distribution of the thermal electrons for the Comptonization \citep{Zdziarski2020}. We also added a Gaussian component to model the emission feature. The model \texttt{tbabs*ThComp*(kerrbb+gaus)} provided the best fit and yielded physically acceptable spectral parameters for the spectra from epochs 6 and 7, only. We tied the spin, inclination, distance, spectral hardening factor, and mass across the two spectra and left the remaining parameters free to vary across the two spectra. We also set $\eta = 0$ as its value was consistent with zero when left free. As discussed in the next section, there are several caveats in putting a constraint on the BH spin, primarily because of absence of firm estimates of BH mass, disc inclination and distance.

In order to estimate the spin of the BH, we performed spectral fitting by varying the mass, distance, inclination and spectral hardening factor. The results are shown in Figure~\ref{fig:mafc}. For these spectral fits, the BH mass was taken in the range 2.5--10 M$_{\sun}$ and the spectral hardening factor was varied in the range 1.5--1.9. This choice of mass range is based on the mass and distance limits proposed by \citet{Nowak2008,Nowak2012}.

\section{Discussion and conclusions}

We report the results from the broad-band spectral analysis of persistent LMXB and BHC 4U 1957+11, by using the \emph{NuSTAR} observations. The source was observed nine times between 2018 September and 2019 November. The X-ray emission evolved during the observations (Figure~\ref{fig:longlc}). The average emission flux was relatively less during the first three epochs, as compared to the subsequent epochs. The emission flux decayed back to the initial level during the last epoch. The source was detected above 20 keV during all the epochs except the first and the last one.

We have used the simple models comprising a multi-temperature disc blackbody for the soft emission, and a power law or Comptonization component for the hard emission along with a Gaussian component to model the spectra. The spectra from epochs 1, 3, 8, and 9 have been modelled by using a power law for the hard emission, while a thermal Comptonization component described the spectra well for epochs 2, 4, 5, 6, and 7.

The disc temperature gradually increased from the first epoch to the seventh epoch (from about 1.36 keV to about 1.79 keV), except for the epoch 3 where it decreased marginally. During the last two epochs, it reduced and reached about 1.47 keV. Similar high disc temperatures have been reported for the BHCs XTE J1748-288 \citep{Miller2001} and LMC X-3 \citep{Wilms2001}. The hard components, powerlaw/Comptonization, are consistent with steep photon indices between 2.22 and 3.32, except for the last epoch where the hard component is not statistically favoured. The high value of the photon indices is indicative of the soft spectral state of the source. LMC X-3 is also known to exhibit a similar simple spectrum during its soft spectral state \citep{Kubota2010}.

The disc normalization for epochs 6 and 7 was observed to be much less ($\sim$ 3.35) as compared to the other relatively low-flux states ($\sim$ 8.12). \citet{Wijnands2002} also reported a decrease in the disc normalization with increasing temperature during their two observations for their spectral analysis with \emph{RXTE}. Similar to the results of \citet{Nowak2008}, these two observations also show increased contribution from the non-thermal component ($\approx 75$ per cent) to the total flux and mimics similarity with their observations identified as the steep power law state.

We have also detected a broad Fe emission line feature in the spectra of the source from all the observations. We have modelled the prominent broad emission feature with a Gaussian component. The upper limits on the line energy are fairly consistent across all the spectra. The large value of the line width is also in agreement with the previous findings \citep{White1986,Singh1994} and can be attributed to the relativistic broadening of the line in the inner region of the disc \citep{Fabian1989}. While this feature is present during all the observations, the contribution of the emission feature remains $< 1$ per cent to the total flux. 

The spectral results presented here are consistent with the previously reported results, especially the trend of the hardening of the spectra with increasing flux \citep{Yaqoob1993,Wijnands2002,Nowak2008}. A similar hardening of the spectra with the increasing flux has been observed for the LMC X-3 \citep{Wilms2001}. We also confirm that the source shows no significant variability on the scale of $2^{-10}$--512 Hz in the 3--20 keV energy band (Figure~\ref{fig:pds}).

We have used the \texttt{kerrbb} component to model the spectra from epochs 6 and 7 to estimate the BH spin and to study its dependence on the mass of BH, its inclination, distance and the spectral hardening factor. For the spectral fits of Figure~\ref{fig:mafc} (a) and (b), we fixed the distance at 7 kpc \citep{Maccarone2020} and inclination at $75^\circ$ \citep{Nowak2008}. The model normalization was set to unity. From Figure~\ref{fig:mafc} (a) and (b), we infer that low spin values occur for a combination of very low mass and spectral hardening factor greater than the theoretically preferred value of 1.7. However, this mass range lies in the known mass gap of the BH and NS distribution \citep{Farr2011,Gomez2015}. A moderate spin of about 0.85 is preferred for BH masses between 4 and 6 M$_{\sun}$. An anti-correlation is observed between the spin and spectral hardening factor. The value of spin decreases as spectral hardening factor increases for a given mass. For masses above 6 M$_{\sun}$, the spin parameter becomes independent of spectral hardening factor and approaches the maximum value 1. The accretion rates follow opposite trend and decrease with increasing BH mass showing an order of magnitude change across the entire mass range. The accretion rate and the spectral hardening factor are correlated, implying a consistent relation with the Suzaku observations \citep{Nowak2012}. Since the distance and BH mass are poorly known, therefore, at this point, it is difficult and beyond the scope of this work, to comment on the scaling relation and the extent of fractional Eddington luminosity radiated from the system. For the \texttt{ThComp} parameters, the best fit results were consistent with photon indices in the range 2.7--2.9 and scattering fraction between 0.7--1.0 for the two spectra.


As a function of distance, the spin and spectral hardening factor are found to be anti-correlated for the mass range of 2.5--10 M$_{\sun}$ (Figure~\ref{fig:mafc} (c), (d)). Our simulations indicate a higher spin for smaller distances. For high mass range ($\sim$ 10 M$_{\sun}$) and at small distances ($<$ 7 kpc), the spin is found to be maximum and almost independent of the distance. As a function of inclination angle (Figure~\ref{fig:mafc} (e), (f)), the spectral fit was poor for higher mass range and at lower disc inclination ($< 70^\circ$). An inverse correlation was observed between the spin and the spectral hardening factor for higher inclination angles ($> 70^\circ$).

The results from \texttt{kerrbb} are consistent with a low spin for low BH masses and high spin for higher masses. Similar high spin values have been found by the previous analysis with the high inclination angles ($\sim 75^\circ$) for low values of the spectral hardening factor \citep{Nowak2008,Nowak2012,Maitra2014}. The high spin value combined with the low mass reconciles with the observed high disc temperature (1.35--1.86 keV) and lower normalization for the source, not usual for other BHC \citep{Nowak2008}.

The spectral properties and X-ray variability of 4U 1954+11 find similarities with other BHCs in the soft spectral state (Swift J1753.5-0127 \citep{Yoshikawa2015}, MAXI J1634-479 \citep{Xu2020} and MAXI J1535-571 \citep{Cuneo2020}). The evolution of the hard component to dominate the emission at the relatively high-flux states as observed during the epochs 6 and 7 provides strong evidence that it does not harbour a NS \citep{Disalvo2000,Wijnands2002}. 

The results presented in this spectral analysis by using the \emph{NuSTAR} observations are consistent with the previous results and provide estimate for the spin of the BH and its possible variation with the primary mass, disc inclination, distance, and spectral hardening factor.  A strong and conclusive result will be possible if the uncertainties in the distance, inclination, and BH mass are resolved.

\normalem
\begin{acknowledgements}

This work has made the use of data obtained with the space mission \emph{NuSTAR}, a project led by the California Institute of Technology, and funded by the NASA. We have used the data archived by the High Energy Astrophysics Science Archive Research Center (HEASARC) online service maintained by the Goddard Space Flight Center. This work has made use of the \emph{NuSTAR} Data Analysis Software (NuSTARDAS) jointly developed by the ASI Space Science Data Center (SSDC, Italy) and the California Institute of Technology (Caltech, USA). RS is supported by the grant awarded to Dr. A. Beri through the INSPIRE faculty award (DST/Inspire/04/2018/001265) by the Department of Science and Technology, Government of India. PS acknowledges the financial support from the Council of Scientific \& Industrial Research (CSIR) under the Junior Research Fellowship (JRF) scheme. 


\end{acknowledgements}
  
\bibliographystyle{raa}
\bibliography{ms2021-0006}

\end{document}